\documentclass[11pt]{article}
\usepackage{amsmath, amssymb, amsfonts, amsthm}
\usepackage{graphicx}
\usepackage{hyperref}

\newcommand{\myendlemmaproof}{\hfill$\triangledown$\vspace{1ex}}
\newcommand{\signw}{w}
\newcommand{\hati}{\hat{\i}}
\newcommand{\hatj}{\hat{\j}}
\newcommand{\hatk}{\hat{\mathrm{k}}}
\newcommand{\sign}{\operatorname{sign}}

\newtheorem{theorem}{Theorem}[section]
\newtheorem{lemma}[theorem]{Lemma}
\newtheorem{proposition}[theorem]{Proposition}
\newtheorem{corollary}[theorem]{Corollary}
\theoremstyle{definition}

\theoremstyle{remark}
\newtheorem{remark}[theorem]{Remark}

\begin{document}

\title{Three-Point Vortex Dynamics as a Lie-Poisson System}
\author{Antonio Hern\'andez-Gardu\~no \\[1ex]
Division of Engineering\footnote{%
The author acknowledges the hospitality of the Mathematics Department at ITAM, where he was a visiting professor while preparing this manuscript.
} \\
Instituto Tecnol\'ogico Aut\'onomo de M\'exico (ITAM) \\
{\footnotesize \texttt{antonio.hernandez.garduno@itam.mx}}\\
}

\date{\today}
\maketitle

\begin{abstract}
This paper studies the reduced dynamics of the three-vortex problem from the point of view of Lie-Poisson reduction on the dual of the Lie algebra of $ U(2) $.  The algebraic study leading to this point of view has been given by Borisov and Lebedev \cite{BoLe98ii, BoLe98iii} (see also \cite{BoBoMa99}).  The main contribution of this paper is to properly describe the dynamics as a Lie-Poisson reduced system on $ (\mathfrak{u}(2) ^\ast, \{\;,\,\}_{\text{LP}} ) $, giving a systematic construction of a one-parameter family of covectors $ \{ \sigma _1, \sigma _2, \sigma _3 \} $ closely related to Pauli spin matrices, and to bring light to the relation between Lie-Poisson reduction and symplectic reduction using Jacobi-Bertrand-Haretu coordinates.
	
\end{abstract}

\tableofcontents

\section{Introduction}

The \emph{Lie-Poisson bracket} $ \{f, h\} _{\pm} (\mu) = \pm \left\langle \mu, \left[ \frac{ \delta f }{ \delta \mu }, \frac{ \delta h }{ \delta \mu } \right] \right\rangle $ defines a natural Poisson structure on the dual $\mathfrak{g}^{\ast}$ of any Lie algebra.  Given a hamiltonian $ h \in \mathcal{F} (\mathfrak{g}^{\ast}) $, a hamiltonian vector field is induced by $ X _h (f) = \{f, h\} _{\pm} $.  The resulting hamiltonian dynamical system is called a \emph{Lie-Poisson} system.
Lie-Poisson system arise naturally through reduction of hamiltonian systems on $ T ^\ast G $, with $ \mathfrak{g} $ the Lie algebra of the Lie group $G$; the reduction of the free rigid body being a prototypical example.  They also arise through Poisson reduction using invariants.  This means that we start with a hamiltonian system defined on a Poisson manifold whose structure is invariant under the action of a Lie group $G$.  In the case when the set $\mathcal{I}$ of $G$-invariants is closed under the Poisson bracket and the result is linear in these invariants then a Poisson structure is induced on $\mathcal{I}$.  For the 3-vortex problem this structure was, to our knowledge, first identified in \cite{BoPa98, BoLe98ii, BoLe98iii}.  For the $N$-body problem, a Lie-Poisson structure on the set of invariants has been studied in \cite{CuBa1997} (for $ N = 2 $) and \cite{LeMoSj1993, BoBoMa99, Du2013}.

This paper deals with the Lie-Poisson system obtained by Poisson reduction of the 3-vortex problem using $ SE(2) $-invariants.   
Generally speaking, the $N$-point-vortex problem arises as a model on incompressible, homogeneous, inviscid fluid flows, governed by Euler's equation on $ \mathbb{R}^2 $, where the vorticity is assumed to be concentrated at $N$ discrete points.  The equations or motion are (see \cite{Newton2001})
\begin{displaymath}
	\dot{ z} _\alpha = \frac{ i }{ 2 \pi} \sum _{ \beta \neq \alpha}^N \Gamma _\beta \frac{ z _\alpha - z _\beta }{ |z _\alpha - z _\beta | ^2 }
\end{displaymath}
where the $ \Gamma _\beta $'s are the vortex strengths.  These equations are equivalent to Hamilton's equations $ \mathbf{ i} _{ X _h } \Omega = d h $ with Hamiltonian
  $$ h = - \frac{ 1 }{ 2 \pi } \sum _{ \alpha < \beta} \Gamma _\alpha \Gamma _\beta \ln | z _\alpha - z _\beta | \;, $$
and symplectic form
\begin{equation} \label{vortexSymplecticForm}
	\Omega _0 (z, w) = - \operatorname{ Im} \sum _{ \alpha = 1} ^n \Gamma _\alpha z _\alpha \bar{ w} _\alpha \;. 
\end{equation}

The Hamiltonian and symplectic form are invariant with respect to the diagonal action of $ SE(2) $ on the phase space of the system identified  with $\mathbb{C}^N$ minus collision points:
\[
  z _i \mapsto e ^{ i \theta} z _i + a \,, \qquad
  (\theta, a) \in SE(2) \cong S ^1 \times \mathbb{C} \,. 
\]
The model admits various conserved quantities related to $ SE(2) $ invariance, time-translation and rescaling symmetries:
\begin{align}\label{eq:circonstants}
  Z _0&= \Gamma_{\text{tot}} ^{-1} \sum _k \Gamma _k z _k \,, \quad I _2  = \frac12 \sum _k \Gamma _k | z _k | ^2 \,,
\end{align}
\[
  \Psi _0= - \sum _{ n<k} \Gamma _n \Gamma _k \ln | z _n - z _k | \,,
\]
\begin{align}
  V_0 &= \frac1{2i} \sum _k \Gamma _k (\bar{z} _k \dot{z} _k - z _k \dot{ \bar{z}} _k ) = \sum _{ n < k} \Gamma _n \Gamma _k \,, \label{eq:virial}
\end{align}
where \( \Gamma_{\text{tot}} := \sum_{k} \Gamma_k \).  The expression
\begin{equation}\label{constantM}
	M = \frac{1}{2 \Gamma _{\text{tot}}} \sum _{ n < k} \Gamma _n \Gamma _k | z _n - z _k | ^2 \,, 
\end{equation}
is also a conserved quantity, but is not independent from the ones mentioned above.  Indeed, it is expressed in terms of $ I _2 $ and $ Z _0 $:
\[
  M = I _2 - \frac{\Gamma_{\text{tot}}}{2} | Z _0 | ^2 \,.
\]

\begin{remark} \label{rem:angularImpulse}
In the point vortex literature, the expression $ \Theta _0 \stackrel{\text{\tiny\rm def}}{=} 2 I _2 $ is referred to as the \emph{angular impulse}.  Also, when $ Z _0 $ is at the origin, it is easy to show that $ J \stackrel{\text{\tiny\rm def}}{=} - I _2 $ is the momentum map for the standard $ SO(2) $ action on $ \mathbb{C} ^N $.  Our notation $ I _2 $ is motivated by a coordinate transformation discussed in section \ref{sec:jbhCoordinates}.
\end{remark}

The symplectic form \eqref{vortexSymplecticForm} induces a Poisson structure on $ \mathbb{C} ^N $.  This Poisson structure can be written in terms of the following group invariants:  the square of the distances between each pair of vortices and the oriented areas of each triad of vortices.  By regarding these quantities as independent, Bolsinov, Borisov and Mamaev \cite{BoBoMa99} are able to describe the reduced vortex dynamics as a subsystem of a hamiltonian system on the Lie algebra $ \mathfrak{u}(n-1) $.  More concretely, Borisov and Lebedev \cite{BoLe98ii, BoLe98iii} study the 3-vortex compact and non-compact vortex dynamics using this point of view.

In this paper we concentrate on the 3-point vortex problem.  We also assume that the total vortex strength is not zero.  One of the objectives is to clarify that, properly understood, the point of view of Borisov and Lebedev lead to a description of the reduced dynamics as a Lie-Poisson reduced system in $ \mathfrak{u} ^\ast (2) $.  The Poisson bracket for the reduced dynamics turns out to be the standard Lie-Poisson braquet on the dual of a Lie algebra (as in \cite[chap. 13]{MaRa99}).

Moreover, this paper contributes by giving a systematic construction of a one-parameter family of covectors $ ( \sigma _1, \sigma _2, \sigma _3 ) $ satisfying either Pauli-commutation relations or relations closely related to them.  The origin of the Pauli symbols is carefully derived using elementary linear algebra tools.  Simple expressions for the Casimirs are then given in terms of coordinates $ (a _0, a _1, a _2, a _3) $ induced by the dual basis of the Pauli symbols.  This allows for the foliation of $ \mathfrak{u} (2) ^\ast $ by level sets of Casimirs be made explicit.  We also relate such construction with an alternate symplectic reduction, namely using canonical transformations involving Jacobi-Bertrand-Haretu coordinates for the original three point vortex system.  (This symplectic reduction is used elsewhere to compute the reconstruction phases of the 3-vortex problem; see \cite{HeSh2018}.)

\section{Extended vortex configuration space}

Consider the Poisson structure on $ \mathbb{C} ^3 $ induced by the symplectic form $ \Omega _0 $:
\[
  \{ f, g \} _{ \mathbb{C} ^3 } = \Omega _0 (X _f, X _g)
\]
with $ X _h $ defined by Hamilton's equation $ i _{ X _h } \Omega _0 = d f $.  Let $ b _i = | z _j - z _k | ^2 $ and let $ \Delta = \Im[ (\overline{z _3 - z _1}) (z _1 - z _2) ] / 2 $, the oriented area of the triangle with vertices at $ z _1, z _2, z _3 $.  Let $ C _3 $ denote the \emph{even permutation 3-cycles of indices $(1,2,3)$}.  It is verified (see \cite{BoLe98ii}) that, with $ (i,j,k) \in C _3 $,
\begin{equation} \label{PoissonStructure}
\begin{split}
  \{ b _i, \Delta \} _{ \mathbb{C} ^3 } &= \frac12 \left[ \left( \frac1{ \Gamma _j } - \frac1{ \Gamma _k } \right) b _i + \left( \frac1{\Gamma _j} + \frac1{\Gamma _k} \right) (b _j - b _k) \right] \,, \\[1ex]
  \{ b _i, b _j \} _{ \mathbb{C} ^3 } &= - \frac{ 8 \, \Delta }{ \Gamma _k } \,. 
\end{split}
\end{equation}

Let $ \mathcal{V} := \mathbb{R} ^4 $ and let $ ( \bar{b} _1, \bar{b} _2, \bar{b} _3, \bar{\Delta} ) $ be the dual to the standard basis in $\mathcal{V}$ (that is to say, the standard projection functionals).  Give $ \mathcal{V} $ a Poisson structure by defining $ \{ \bar{b} _i, \bar{\Delta} \} _{\mathcal{V}} $ and $ \{ \bar{b} _i, \bar{b} _j \} _{\mathcal{V}} $ as the right-hand sides of \eqref{PoissonStructure}, putting bars on top of the $ b _i $'s and $\Delta$.  Then $ \psi : \mathbb{C} ^3 \longrightarrow \mathcal{V} $ given by $ (z _1, z _2, z _3) \mapsto (b _1, b _2, b _3, \Delta) $ is an $ SE(2)$-invariant Poisson map.  Moreover, restriction of $ \psi $ to $ P _0 \subset \mathbb{C} ^3 $, where $ P _0 \subset \mathbb{C} ^3 $ is the set of vortex configurations with center of vorticity at the origin, gives an $ SO(2) $-invariant Poisson map.

We call $\mathcal{V}$ the \emph{extended vortex configuration space}.  It is to be regarded as the space of ``triangle configurations'' with sides of length $ \sqrt{b _i} $ and oriented area $\Delta$.  Of course, only those points in $\mathcal{V}$ for which $ b _i \ge 0 $ and satisfy Heron's condition relating the area and the sides of a triangle,
\begin{equation} \label{HeronCondition}
	(4 \Delta) ^2 + b _1 ^2 + b _2 ^2 + b _3 ^2 - 2(b _1 b _2 + b _2 b _3 + b _3 b _1) = 0 \,, 
\end{equation}
have physical meaning.

Let $ H(b _1, b _2, b _3, \Delta) $ be defined as the left-hand side of \eqref{HeronCondition}.  We will refer to $H$ as \emph{Heron's function}.  It is verified that $H$ is a Casimir of the Poisson structure $ \{ \;,\, \} _{\mathcal{V}} $.

\subsection{Casimirs}

Observe that $ \{ \;,\, \} _{\mathcal{V}} $ is closed in $ \mathcal{V} ^\ast \subset \mathcal{F} (\mathcal{V}) $; that is to say, the braquet of two linear functionals is again a linear functional.  It follows that $ \{ \;,\, \} _\mathcal{V} $ makes $ \mathcal{V} ^\ast $ a four-dimensional real Lie algebra.  It's center $ Z(\mathcal{V} ^\ast) $ is
\[
  Z(\mathcal{V} ^\ast) = \operatorname{span}(\sigma _0) \,, \quad \text{where} \quad \sigma _0 := \frac1{2 \Gamma_{\text{\rm tot}}} \sum _{(i,j,k) \in C _3}  \Gamma _j \Gamma _k \, \bar{b} _i \,. 
\]
As a consequence, 

\begin{proposition} \label{prop:MCasimir}
$M$, as defined in \eqref{constantM}, is a Casimir of $ \{ \;,\, \} _{\mathcal{V}} $.
\end{proposition}

Besides $M$, we also have that

\begin{proposition}
$H$ is a Casimir of $ \{ \;,\, \} _{\mathcal{V}} $.
\end{proposition}

\begin{proof}
It is easily verified that 
\[
	\frac{\partial H }{ \partial b _1 } \{ \bar{b} _1, \xi \} + \frac{\partial H }{ \partial b _2 } \{ \bar{b} _2, \xi \} + \frac{\partial H }{ \partial b _3 } \{ \bar{b} _3, \xi \} + \frac{\partial H }{ \Delta } \{ \bar{\Delta}, \xi \} = 0
\]
for $ \xi = \bar{b} _1, \bar{b} _2, \bar{b} _3, \bar{\Delta} $.
\end{proof}

\section{Splitting of vortex algebra}

Let $ \mathcal{W} ^\ast = \operatorname{span} \{ \bar{b} _1, \bar{b} _2 , \bar{b} _3 \} \subset \mathcal{V} ^\ast  $ and consider the linear transformation
\[
	A : \mathcal{W} ^\ast \longrightarrow \mathcal{W} ^\ast \,, \quad x \mapsto \{ x, \bar{\Delta} \} _{ \mathcal{V} } \,. 
\] 
A convenient basis of $ \mathcal{W} ^\ast $ to work with is $ \beta = \{ \bar{b} _1 / \Gamma _1 , \bar{b} _2 / \Gamma _2, \bar{b} _3 / \Gamma _3 \} $.  The matrix representation of $A$ is then given by
\begin{equation} \label{aaBeta}
  [A] _\beta = \frac1{2 \, \Gamma _1 \Gamma _2 \Gamma _3}
		\begin{pmatrix}
			\Gamma _1 (\Gamma _3 - \Gamma _2) & - \Gamma _1 (\Gamma _1 + \Gamma _3) & \Gamma _1 (\Gamma _1 + \Gamma _2) \\[1ex]
			\Gamma _2 (\Gamma _2 + \Gamma _3) & \Gamma _2(\Gamma _1 - \Gamma _3) & -\Gamma _2 (\Gamma _1 + \Gamma _2) \\[1ex]
			-\Gamma _3 (\Gamma _2 + \Gamma_3) & \Gamma _3 (\Gamma _1 + \Gamma _3) & \Gamma _3 (\Gamma _2 - \Gamma _1)
		\end{pmatrix} \,. 
\end{equation}
The eigenvalues of $A$ are
\[
	\operatorname{spectrum}(A) = \left\{ 0, -i \sqrt{W _0}, i \sqrt{W _0} \right\}
\] 
where 
\[
	W _0 \stackrel{\text{\tiny\rm def}}{=} \frac1{\Gamma _1 \Gamma _2} + \frac1{\Gamma _2 \Gamma _3} + \frac1{\Gamma _3 \Gamma _1} \,. 
\] 
Since we are assuming $ \Gamma _{\text{tot}} \neq 0 $,we have $ W _0 \neq 0 $.  Hence $ \ker A = Z(\mathcal{V} ^\ast) = \operatorname{span}(\sigma _0) $.  Note that $ [\sigma _0] _\beta = (1,1,1) / (2 W _0) $.

Let $ \mathcal{S} = \operatorname{range} (A) \subset \mathcal{W} ^\ast $.  It is verified that $ \sigma _0 \not\in \mathcal{S} $.  Therefore,
\begin{proposition} \label{prop:isomorphismS}
	Let $ A : \mathcal{W} ^\ast \longrightarrow\mathcal{W}^\ast $ be defined by $ A(x) = \{ x, \bar{\Delta} \} _{ \mathcal{V} } $.  Then,
	\begin{enumerate}
	\item $ \mathcal{W} ^\ast = \mathcal{S} \oplus Z(\mathcal{V} ^\ast) $.
	\item $ \left. A \right| _{ \mathcal{S} } $ is an isomorphism.
	\item $ \left. A ^2 \right| _{ \mathcal{S} } = - W _0 \operatorname{Id} $.
	\end{enumerate}
(Here ``$ \operatorname{Id} $'' denotes the identity.)
\end{proposition}
\begin{proof}
The first two claims follow directly from the definition of $ \mathcal{S} $ and the fact that $ \sigma _0 \not\in  \mathcal{S} $.  The last claim follows from the spectrum of $A$.
\end{proof}
Let us now consider the bilinear form on $\mathcal{S}$ given by $ (x,y) \mapsto \{x, \{y, \bar{\Delta} \}\} $.  It is easy to check that this is symmetric.  Let $ Q : \mathcal{S} \longrightarrow\mathbb{R} $ be its associated quadratic form, i.e.
\[
	Q(x) \bar{\Delta} \stackrel{\text{\tiny\rm def}}{=} \{ x, \{x, \bar{\Delta} \} \} \,. 
\]

\begin{proposition} \label{prop:QNegDef}
	$Q$ is negative-definite if $ W _0 > 0 $ and indefinite if $ W _0 < 0 $.
\end{proposition}

\begin{proof}
	Let $  v_i \in \mathcal{W} ^\ast $ such that its representation in basis $\beta$ is the $i$-th column of 
	\[
	  \mathbf{V} = \begin{bmatrix} 0 & -\gamma _3 & \gamma_2 \\ \gamma _3 & 0 & -\gamma _1 \\ -\gamma_2 & \gamma _1 & 0 \end{bmatrix} \,, \quad
	\gamma _i := \Gamma _j + \Gamma _k \,, \quad (i,j,k) \in C _3 \,. 
	\] 
	A computation shows that
	\[
	  \mathbf{V} = [A] _\beta \begin{bmatrix} 0 & \gamma _3 & \gamma _2 \\ \gamma _3 & 0 & \gamma_1 \\ \gamma _2 & \gamma _1 & 0 \end{bmatrix}
	\begin{bmatrix} \Gamma _1 && \\ & \Gamma _2 & \\ && \Gamma _3 \end{bmatrix} \,. 
	\] 
	Hence $ v _i \in \operatorname{range} A = \mathcal{S} $, $ i = 1, 2, 3 $.  The condition $ \Gamma _{\text{tot}} \neq 0 $ implies that there is $ \{a, b\} \subset \{1, 2, 3 \} $ such that $ \{ v _a, v _b \} $ is a basis of $S$.  It is easy to verify that
	\begin{equation} \label{quadraticvavb}
	  \mathbb{Q}_{ ab } \stackrel{\text{\tiny\rm def}}{=} [Q] _{ (v _a, v _b) } = - \frac{ 16 W _0 \gamma _c }{ \Gamma _c } \begin{bmatrix} \gamma _b / \Gamma _b & -1 \\ -1 & \gamma _a / \Gamma _a \end{bmatrix} \,, 
	\end{equation}
	with $ (a, b, c) \in C _3 $, and
	\[
	  \det \mathbb{Q} _{ ab } = 256 \gamma _c ^2 W _0 ^3 \,, \quad (a, b, c) \in C _3 \,. 
	\] 
	Note that $ \gamma _c \neq  0 $, for otherwise $ v _a $ and $ v _b $ would be parallel.  Hence $Q$ is definite if $ W _0 > 0 $ and indefinite if $ W _0 < 0 $.
	
	Observe that the case $ W _0 < 0 $ contains the case when not every pair in $ \{ v _1, v _2, v _3 \} $ is a basis of $S$.  For if $ \{ v _a, v _b \} $ is basis of $\mathcal{S}$ and $ v _c \parallel v _a $ ($ v _c \parallel v _b$) then $ \gamma _b = 0 $ ($ \gamma _a = 0 $), thus implying $ W _0 = -1/ \Gamma _a ^2 $ ($ W _0 = - 1/ \Gamma _b ^2 $).
	
	Now, consider the case $ W _0 > 0 $, i.e. $Q$ is definite.  Then the entries in the main diagonal of $ \mathbb{Q} _{ ab } $ have the same sign, for all three cases $ (a,b) = (1,2), (2,3) $ and $ (3,1) $.  That is to say, $ \sign \gamma _1 / \Gamma _1 = \sign \gamma _2 / \Gamma _2 = \sign \gamma _3 / \Gamma _3 $.  Thus, from the expression of $ \mathbb{Q} _{ ab } $ above, $Q$ is negative definite.
\end{proof}

\begin{remark}
	The basis $ (v _a, v _b) $ of $ \mathcal{S} $, $ \{a, b\} \subset \{1, 2, 3 \} $, constructed in the proof of proposition \ref{prop:QNegDef} will be used in various places below.  Note that, referring to the notation in the proof, $ V \mathbf{x} = (\gamma _1, \gamma _2, \gamma _3) \times \mathbf{x} $ for all $ \mathbf{x} \in \mathbb{R} ^3 $.  Hence $ (\gamma _1, \gamma _2, \gamma _3) \in \ker V $ and 
	\begin{equation} \label{crossv0}
	  [x] _\beta \times (\Gamma _2 + \Gamma _3, \Gamma _3 + \Gamma _1, \Gamma _1 + \Gamma _2 ) = 0 \quad \text{for all $ x \in \mathcal{S} $} \,,
	\end{equation}
	which is another characterization of $\mathcal{S}$.
\end{remark}

For further use, let us state a useful matrix representation of $Q$ restricted to $\mathcal{S}$ with respect to basis $\beta$.

\begin{proposition} \label{prop:qqM}
	Let $ \tilde{Q} : \mathcal{W} ^\ast \longrightarrow \mathbb{R} $ be the quadratic form whose matrix representation w.r.t. $\beta$ is:
	\[
		[\tilde{Q}] _\beta = - \frac8{\Gamma _1 ^2 \Gamma _2 ^2 \Gamma _3 ^3} \, \mathbb{Q} \,,
	\]
	with
	\[
		\mathbb{Q} \stackrel{\text{\tiny\rm def}}{=} 
		\begin{pmatrix}
			(\Gamma _2 + \Gamma _3) ^2 & \Gamma _1 \Gamma _2 & \Gamma _1 \Gamma _3 \\
			\Gamma _1 \Gamma _2 & (\Gamma _3 +\Gamma _1) ^2 & \Gamma _2 \Gamma _3 \\
			\Gamma _1 \Gamma _3 & \Gamma _2 \Gamma _3 & (\Gamma _1 + \Gamma _2) ^2 
		\end{pmatrix} \,.
	\]
Then $ Q \big| _{ \mathcal{S} } = \tilde{Q} \big| _{ \mathcal{S} } $.
\end{proposition}
\begin{proof}
It suffices to verify that the bilinear forms associated with $ Q $ and $ \tilde{Q} $ coincide when evaluated at basis elements $ v _a, v _b $.
\end{proof}

\section{Classification of vortex algebra}

Having defined a quadratic form on $\mathcal{S}$, we now consider its level set for a negative value.  We will now see that to each point in this level set we associate a basis of $ \mathcal{V} ^\ast $ whose elements we will call ``Pauli-symbols'',  since their commutation relations are closely related to those of the standard Pauli matrices.

\subsection{Pauli symbols}
\label{sec:paulisymbols}

Let 
\[
	\tilde{ \mathcal{S} } \stackrel{\text{\tiny\rm def}}{=} \{ x \in \mathcal{S} \mid Q(x) = -1 \} \,. 
\] 
Note that, topologically, $ \tilde{ \mathcal{S} } $ is a circle if $ W _0 > 0 $ and an open interval if $ W _0 < 0 $.

\begin{theorem}
	Given $ x \in \tilde{\mathcal{S}} $,  let
	\begin{equation} \label{pauliSymbolsDefined}
	  \sigma _1 = \frac{ 2 }{ \sqrt{| W _0 |} } \{ x, \bar{\Delta} \} \,, \quad
	  \sigma _2 = \frac{ 2 }{ \sqrt{| W _0 |} } \bar{\Delta} \,, \quad
	  \sigma _3 = 2x \,. 
	\end{equation}
	Then
	\begin{equation} \label{vortexPauliCommRel}
	  \{ \sigma _1, \sigma _2 \} = -2 w \sigma _1 \,, \quad
	  \{ \sigma _2, \sigma _3 \} = -2 \sigma _1 \,, \quad
	  \{ \sigma _3, \sigma _1 \} = - 2 \sigma_2 
	\end{equation}
	where $ w = \sign W _0 $.
\end{theorem}

\begin{proof}
	Let $ j = 1 $ if $ W _0 > 0 $, and $ j = i $ if $ W _0 < 0 $.  Then
	$ \sigma _1 = 2j \{ x, \bar{\Delta} \} / \sqrt{W _0} $ 
	and
	$ \sigma _2 = 2j \bar{\Delta} / \sqrt{W _0} $.  We compute:
	\[
	  \{ \sigma _1, \sigma _2 \} = \frac{ 4 j ^2 }{ W _0 } \{ \{ x, \bar{\Delta} \}, \bar{\Delta} \} = \frac{ 4 j ^2 }{ W _0 } A ^2 (x) = -2 \sign(W _0) \sigma_3 \,, 
	\] 
	using proposition \ref{prop:isomorphismS} in the last equality;
	\[
	  \{ \sigma _2, \sigma _3 \} = \frac{ 4j }{ \sqrt{W _0} } \{ \bar{\Delta}, x \} = - 2 \sigma _1 \,;
	\] 
	\[
	  \{ \sigma _3, \sigma _1 \} = \frac{ 4j }{ \sqrt{W _0} } \{ x, \{x, \bar{\Delta} \} \} = \frac{ 4j }{ \sqrt{W _0} } Q(x) \bar{\Delta} = - 2 \sigma _2 \,. 
	\] 
\end{proof}

As a consequence of the theorem we can identify the Lie algebra associated with $ \mathcal{V} ^\ast $:

\begin{corollary} \label{cor:firstLieAlgebraCharacterization}
	\[
	  \mathcal{S} \oplus \operatorname{span}(\bar{\Delta}) \cong \left\{
		\begin{array}{r@{\hspace{1ex}\text{if}\hspace{1ex}}l}
			\mathfrak{su}(2) & W _0 > 0 \,, \\[1ex]
			\mathfrak{su}(1,1) & W _0 < 0 \,. 
		\end{array} \right.
	\] 
\end{corollary}

\begin{proof}
	From proposition \ref{prop:isomorphismS} we get that $ (\sigma _1, \sigma _2) $ is a basis of $\mathcal{S}$.  When $ W _0 > 0 $, prescribe the identification $ \sigma _k \mapsto i \tilde{\sigma} _k $, $ k = 1,2,3 $, where
	\[
	  \tilde{\sigma} _1 = \begin{bmatrix} 0 & 1 \\ 1 & 0 \end{bmatrix} \,, \quad
	  \tilde{\sigma} _2 = \begin{bmatrix} 0 & -i \\ i & 0 \end{bmatrix} \,, \quad
	  \tilde{\sigma} _3 = \begin{bmatrix} 1 & 0 \\ 0 & -1 \end{bmatrix} 
	\] 
	are the Pauli spin matrices.  Then \eqref{vortexPauliCommRel} transforms into the standard Pauli commutation relations.  Since $ (i \tilde{\sigma} _1, i \tilde{\sigma} _2, i \tilde{\sigma} _3) $ is a basis of $ \mathfrak{su}(2) $, the claim follows.
	
	When $ W _0 < 0 $, use the identification $ \sigma _k \mapsto i \tilde{\varsigma} $, $ k = 1, 2, 3 $, where
	\[
	  \tilde{\varsigma} _1 = \begin{bmatrix} 0 & 1 \\ -1 & 0 \end{bmatrix} \,, \quad
	  \tilde{\varsigma} _2 = \begin{bmatrix} 0 & -i \\ -i & 0 \end{bmatrix} \,, \quad 
	  \tilde{\varsigma} _3 = \begin{bmatrix} 1 & 0 \\ 0 & -1 \end{bmatrix}
	\] 
	are ``modified Pauli spin matrices''.  Then \eqref{vortexPauliCommRel} transforms into the commutation relations satisfied by the $ \tilde{\varsigma} _i $'s.  Moreover, $ (i \tilde{\varsigma} _1, i \tilde{\varsigma} _2, i \tilde{\varsigma} _3) $  is a basis of $ \mathfrak{su}(1,1) $.  Hence the claim follows.
\end{proof}

This corollary will allow us to identify $ \mathcal{V} ^\ast $ as a Lie-algebra.  Moreover, the next lemma will allow us to identify $\mathcal{V}$ as a Lie-Poisson system.

\begin{lemma} \label{lem:adjointPoisson}
	Let $ ( V, \{ \;,\, \} ) $ be a Poisson vector space such that $ V ^\ast \subset \mathcal{F} (V) $ is closed under $ \{ \;,\, \} $; hence $ ( V ^\ast, \{ \;,\, \} ) $ is a Lie algebra.  Suppose that $ \psi : V ^\ast \longrightarrow \mathfrak{g} $ is a Lie algebra isomorphism, i.e.
	\[
	  [ \psi(\alpha), \psi(\beta) ] = \psi( \{ \alpha, \beta \} )
	\]
	for all $ \alpha, \beta \in V ^\ast $.  (Here $ [ \;,\, ] $ denotes the Lie bracket on $\mathfrak{g}$.)
Let $ \varphi : \mathfrak{g}^{\ast} \longrightarrow V $ be the adjoint operator to $ \psi $, i.e. defined by
\begin{equation} \label{adjointOperator}
  \left\langle \alpha, \varphi (\mu) \right\rangle = \left\langle \mu, \psi (\alpha) \right\rangle \,. 
\end{equation}
Then $ \varphi $ is a Poisson transformation; that is to say,
\[
  \{ f \circ \varphi, h \circ \varphi \} _{\text{LP}} = \{ f, h \} \circ \varphi 
\]
for all $ f, h \in \mathcal{F} (V) $.  (Here $ \{ \;,\, \} _{\text{LP}} $ denotes the Lie-Poisson bracket on $ \mathfrak{g} ^\ast $; for its definition see \cite[chap. 10]{MaRa99}.)
\end{lemma}
\emph{Proof:}  It suffices to consider the case $ f = \alpha, g = \beta $, 
with $ \alpha, \beta \in V ^\ast $.  Observe that
\[
  (f \circ \varphi) (\mu) = \langle f \circ \varphi, \mu \rangle = \langle \alpha, \varphi(\mu) \rangle = \langle \mu, \psi(\alpha) \rangle \,. 
\]
Thus, under the identification $ \mathfrak{g}  ^{\ast\ast} = \mathfrak{g} $, $ f \circ \varphi = \psi(\alpha) $.  Analogously, $ g \circ \varphi = \psi(\beta) $.  Hence
\[\begin{split}
	\{ f \circ \varphi, g \circ \varphi \} _{\text{LP}} (\mu) &= \left\langle \mu, \left[ \frac{ \delta (f \circ \varphi) }{ \delta \mu }, \frac{ \delta (g \circ \varphi) }{ \delta \mu } \right] \right\rangle \\
	&= \left\langle \mu, \left[ \psi(\alpha), \psi(\beta) \right] \right\rangle = \left\langle \mu, \psi \left( \{ \alpha, \beta \} \right) \right\rangle \\
	&= \left\langle \{ \alpha, \beta \}, \varphi(\mu) \right\rangle = \left\langle \{ f, g \} \circ \varphi, \mu \right\rangle \\
	&= \left( \{f, g \} \circ \varphi \right) (\mu)
\end{split}\] 
as claimed.  $\square$

\begin{theorem} \label{theo:secondLieAlgebraCharacterization}
Let $ \mathcal{V} \cong \mathbb{R} ^4 $ be the extended vortex configuration space.  Then $ \mathcal{V} ^\ast \cong \mathfrak{g} $, where 
\[
	\mathfrak{g} \cong \left\{ \begin{array}{l@{\hspace{0.4em}\text{if}\hspace{0.5em}}l}
		\mathfrak{u} (2) & W _0 > 0 \,, \\
		\mathfrak{u} (1,1) & W _0 < 0 \,. 
	\end{array} \right.
\] 
Moreover, $ \mathcal{V} \cong \mathfrak{g} ^\ast $ is a Lie-Poisson system and its Poisson bracket $ \{\;,\,\}_{ \mathcal{V} } $ is identified with the standard Lie-Poisson bracket $ \{\;,\,\}_{\text{LP}} $ given by  
\[
	\{f, g\}_{\text{LP}} \, (\mu) \stackrel{\text{\tiny\rm def}}{=} \left\langle \mu, \left[ \frac{ \partial f }{ \partial \mu } , \frac{ \partial g }{ \partial \mu } \right] \right\rangle \,. 
\] 
\end{theorem}

\begin{remark}. The Lie-Poisson bracket $ \{\;,\,\}_{\text{LP}} $ is the natural Poisson structure on the dual of a Lie-algebra.  See \cite[chap. 13]{MaRa99}.
\end{remark}

\begin{proof}
Since $ \mathcal{V} ^\ast = \mathcal{S} \oplus \operatorname{span}(\bar{\Delta}) \oplus \operatorname{span}(\sigma _0) $, where $ \sigma _0 $ generates the center of $ \mathcal{V} ^\ast $, the first claim follows at once from corollary \ref{cor:firstLieAlgebraCharacterization}.  The second claim then follows from lemma \ref{lem:adjointPoisson}.
\end{proof}

\subsection{Casimirs in ``Pauli-coordinates''}

The symplectic leaves of $ \mathcal{V} $ are its coadjoint orbits.  These are the common level sets of Casimir functions $M$ and $H$.  (These were defined in \eqref{constantM} and \eqref{HeronCondition}, respectively.). The Pauli symbols just constructed provide privileged coordinates in the sense that expressions for the Casimirs take very simple forms when expressed in these coordinates.

Let $ (\sigma ^0, \sigma ^1, \sigma ^2, \sigma ^3) $ be the basis of $\mathcal{V}$ dual to the basis of $ \mathcal{V} ^\ast $ given by $ (\sigma _0, \sigma _1, \sigma _2, \sigma _3) $.  (Here $ \sigma _k $, $ k = 1, 2, 3 $, are the Pauli symbols defined in \eqref{pauliSymbolsDefined} and $ \sigma _0 $ was defined before proposition \ref{prop:MCasimir}.)  Let $ (a _0, a _1, a _2, a _3) $ be coordinates of $ \mathcal{V} $ with respect to basis $ (\sigma ^0, \sigma ^1, \sigma ^2, \sigma ^3) $.  Note that these coordinates depend on the choice of $ x \in \tilde{ \mathcal{S} } $.

\begin{proposition} \label{prop:casimirsPauliCoords}
For every $ x \in \tilde{ \mathcal{S} } $, the expressions for Casimirs $M$ and $H$ in terms of coordinates $ a _k $ are:
\begin{align}
	M &= a _0 \,, \\[2ex]
	H &= \left\{ \begin{array}{l@{\quad\text{if}\quad}l}
		4 W _0 (- a _0 ^2 + a _1 ^2 + a _2 ^2 + a _3 ^2) & W _0 > 0 \,, \\[2ex]
		4 W _0 (- a _0 ^2 - a _1 ^2 - a _2 ^2 + a _3 ^2) & W _0 < 0 \,. 
	\end{array} \right. \label{shapeConics}
\end{align}
\end{proposition}
	
\begin{proof}
	Let $ \hat{ v } _a = v _a / | Q( v _a ) | ^{ 1/2 } $, hence $ Q( \hat{ v } _\alpha) = \pm 1 $.  Let $ \hati = \hat{ v } _a $ if $ Q(\hat{ v } _\alpha) = -1 $, which is necessarily the case if $ W _0 > 0 $.  Otherwise let $ \hati = \{ \hat{ v } _\alpha, \bar{\Delta} \} / | W _0 | ^{ 1/2 } $.  Furthermore, let $ \hatj = \{ \hati, \bar{\Delta} \} / | W _0 | ^{ 1/2 } $ and $ \hatk = \sigma _0 /2 $.

\begin{lemma} \label{lem:quadraticij}
	$ Q(\hatj) = \signw\, Q(\hati) = - \signw $, where $ \signw = \sign{W _0} $.  Moreover, if $B$ denotes the symmetric bilinear form associated to $Q$, then $ B(\hati, \hatj) = 0 $.
\end{lemma}
\noindent\emph{Proof of lemma.}
It is easy to verify that $$ \{ v _a, \bar{\Delta} \} = \frac1{ \Gamma _c}\, v _a + \frac{\Gamma _a + \Gamma _c}{\Gamma _b \Gamma _c} \, v _b \,.  $$  The claims  then follow from direct computations using the expression for $ \mathbb{Q}_{ab} $ given in \eqref{quadraticvavb}. \myendlemmaproof

Now, let $ \underline{\hati} $ be the column vector representing $ \hati $ in basis $\beta$ and similarly for $ \underline{\hatj} $ and $ \underline{\hatk} $.  Consider the change of basis matrix $ P = [ \underline{\hati} \; \underline{\hatj} \; \underline{\hatk} ] $.  A direct computation shows that

\begin{lemma}\label{lem:repAijk}
The representation of $A$ in basis $ (\hati, \hatj, \hatk) $ is 
\[
	P ^{-1} [A] _\beta P = \sqrt{| W _0 |} \begin{bmatrix} J & 0 \\ 0 & 0 \end{bmatrix}
\]
where 
\[
	J = \left\{ \begin{array}{c@{\quad\text{if}\quad}l}
		\begin{bmatrix} 0 & -1 \\ 1 & 0 \end{bmatrix} & W _0 > 0 \,, \\[2ex]
		\begin{bmatrix} 0 & 1 \\ 1 & 0 \end{bmatrix} & W _0 < 0 \,. 
	\end{array} \right.
\]	
\end{lemma}
\vspace{1ex}

Let us now define a transformation $ \mathcal{R} (\theta) : \mathcal{W} ^\ast \longrightarrow \mathcal{W} ^\ast $ parametrized by $ \theta \in \mathbb{R} $ as follows.  First, let
\[
	R(\theta) = \left\{ \begin{array}{c@{\quad\text{if}\quad}l}
		\begin{pmatrix} \cos \theta & - \sin \theta \\ \sin \theta & \cos \theta \end{pmatrix}
			& W _0 > 0 \,, \\[2ex]
		\begin{pmatrix} \cosh \theta & \sinh \theta \\ \sinh \theta & \cosh \theta \end{pmatrix}
			& W _0 < 0 \,. 
	\end{array} \right.
\] 
Let $ R _{ e _3 } (\theta) = \begin{pmatrix} R(\theta) & 0 \\ 0 & 1 \end{pmatrix} $.  Finally, let $ \mathcal{R} (\theta) $ be defined by
\[
	[ \mathcal{R} (\theta) ] _\beta = \tilde{R}(\theta) \stackrel{\text{\tiny\rm def}}{=} P \, R _{ e _3 } (\theta) \, P ^{-1} \,. 
\] 

\begin{lemma} \label{lem:commutesWithRotation}
	$ \mathcal{R} (\theta) A = A \, \mathcal{R} (\theta) $.
\end{lemma}
\noindent\emph{Proof of lemma.}
Observe that $ [\mathcal{R}(\theta) A] _\beta P = P \, R _{ e _3 } (\theta) (P ^{-1} [A] _\beta P) $.  Using lemma \ref{lem:repAijk} and noting that, for both $ W _0 > 0 $ and $ W _0 < 0 $, we have $ R(\theta) J = J R(\theta) $, it follows that 
\[
	[\mathcal{R}(\theta) A] _\beta P = P (P ^{-1} [A] _\beta P) R _{ e _3 } (\theta) = [A] _\beta P R _{ e _3 } (\theta) \,. 
\] 
Multiplying by $ P ^{-1} $ on the right, $ [\mathcal{R}(\theta) A] _\beta = [ A \, \mathcal{R} (\theta)] _\beta $.  \myendlemmaproof

Let $ \hat{v} = \cos \theta \, \hati + \sin \theta \, \hatj $ if $ W _0 > 0 $, and $ \hat{v} = \cosh \theta \, \hati + \sinh \theta \, \hatj $ if $ W _0 < 0 $.  By lemma \ref{lem:quadraticij}, $ Q(\hat{v}) = -1 $.  It is clear that every $ x \in \tilde{ \mathcal{S} } $ (defined at the beginning of section \ref{sec:paulisymbols}) is equal to $ \hat{v} $ for some $ \theta \in \mathbb{R} $.  Let $ \sigma _k $, $ k = 1, 2, 3 $, be given by \eqref{pauliSymbolsDefined} with $ x = \hat{v} $.
%Let $ \sigma _0 $ be defined by $ [\sigma _0] _\beta = (1,1,1) / (2 W _0) $.
Note that $ \sigma _1 = | W _0 | ^{ -1/2 } A \, \sigma _3 $.  It is easy to see that $ \sigma _3 = \mathcal{R} (\theta) (2 \hati) $ and, using lemma \ref{lem:commutesWithRotation}, that $ \sigma _1 = \mathcal{R} (\theta) (2 \hatj) $.  Moreover, $ \mathcal{R}(\theta) \sigma _0 = \sigma _0 $.  Therefore,

\begin{lemma} \label{lem:rotationPMatrix}
	Let $ s _k = [\sigma _k] _\beta $, $ k = 0, 1, 3 $.  Then $ \tilde{P} \stackrel{\text{\tiny\rm def}}{=} [s _3 \, s _1 \, s _0] = 2 \tilde{R}(\theta) P $.
\end{lemma}

Recall that Heron's function, expressed in coordinates $ (b _1, b _2, b _3, \Delta) $, is given by \eqref{HeronCondition}, so that
\[
	H= (4 \Delta) ^2 + [b _1 \; b _2 \; b _3] \; \mathbb{H} \begin{bmatrix} b _1 \\ b _2 \\ b _3 \end{bmatrix} \,, 
\]
with $ \mathbb{H} = \begin{bmatrix} 1 & -1 & -1 \\ -1 & 1 & -1 \\ -1 & -1 & 1 \end{bmatrix} $.
Let $ \Gamma = \begin{bmatrix} \Gamma _1 \\ & \Gamma _2 \\ && \Gamma _3 \end{bmatrix} $.  The matrix that changes $ (\sigma _3, \sigma _1, \sigma _0, \sigma _2) $-coordinates to $ (\bar{b} _1, \bar{b} _2, \bar{b} _3, \bar{\Delta}) $-coordinates is given by $ \begin{bmatrix} \Gamma ^{-1} \tilde{P} \\ & 2 | W _0 | ^{ -1/2 } \end{bmatrix} $.  Hence, if $ (a _3, a _1, a _0, a _2) $ are coordinates with respect to the dual basis to $ (\sigma _3, \sigma _1, \sigma _0, \sigma _2) $, the change of coordinates matrix taking $ (a _3, a _1, a _0, a _2) $ to $ (b _1, b _2, b _3, \Delta) $ is given by $ \begin{bmatrix} \Gamma \tilde{P} ^{ -T } \\ & | W _0 | ^{ 1/2 } / 2 \end{bmatrix} $.  Therefore, Heron's function expressed in coordinates $ a _k $ is given by
\[
	H = 4 | W _0 | a _2 ^2 + [ a _3 \; a _1 \; a _0 ] \; \mathbb{H} ' \begin{bmatrix} a _3 \\ a _1 \\ a _0 \end{bmatrix} \,,
\] 
where, 
\[
	\mathbb{H} ' = \tilde{P} ^{-1} \Gamma \, \mathbb{H} \, \Gamma \tilde{P} ^{ -T } 
		= \frac14 R _{ e _3 } (-\theta) P ^{-1} \Gamma \, \mathbb{H} \, \Gamma P ^{ -T } R _{ e _3 } (-\theta) ^T \,. 
\]
(The second equality follows from lemma \ref{lem:rotationPMatrix} and $ \tilde{R}(\theta) = P \, R _{ e _3 } (\theta) \, P ^{-1} $.)
A direct computation shows that 
\[
	\mathbb{H} '' \stackrel{\text{\tiny\rm def}}{=} \frac14 P ^{-1} \Gamma \, \mathbb{H} \, \Gamma P ^{ -T } = 4 W _0 \begin{bmatrix} 1 \\ & \signw \\ && -1 \end{bmatrix} \,,
\] 
where $ \signw = \sign W _0 $.  It is easy to check that, for both $ W _0 > 0 $ and $ W _0 < 0 $, $ R(-\theta) \begin{bmatrix} 1 \\ & \signw \end{bmatrix} R(-\theta) ^T = \begin{bmatrix} 1 \\ & \signw \end{bmatrix} $.  Hence $ \mathbb{H} ' = \mathbb{H} '' $, and
\[
	H = 4 W _0 \left( \signw \, a _2 ^2 + [ a _3 \; a _1 \; a _0 ] \begin{bmatrix} 1 \\ &\signw \\ && -1 \end{bmatrix} \begin{bmatrix} a _3 \\ a _1 \\ a _0 \end{bmatrix} \right) \,, 
\] 
which proves the second claim of proposition \ref{prop:casimirsPauliCoords}.

As for the first claim, let $ v = (b _1, b _2, b _3, \Delta) \in \mathcal{V} $.  Note that $ I _2 $ is given by \eqref{constantM}, since $ Z _0 = 0 $.  Thus,
\begin{align*}
	I _2 (v) &= \frac{ \Gamma _1 \Gamma _2 \Gamma _3 }{ 2 \Gamma _{\text{tot}} } \left( \frac{ b _1 }{ \Gamma _1 } + \frac{ b _2 }{ \Gamma _2 } + \frac{ b _3 }{ \Gamma _3 } \right) \\
		&= \frac1{2 W _0} \left[ \frac1{\Gamma _1} \; \frac1{\Gamma _2} \; \frac1{\Gamma _3} \right] \begin{bmatrix} b _1 \\ b _2 \\ b _3 \end{bmatrix} \,. 
\end{align*}
Hence $ [I _2] _{ (\bar{b} _1, \bar{b} _2, \bar{b} _3) } = \frac1{2 W _0} \begin{bmatrix} 1/ \Gamma _1 \\ 1/ \Gamma _2 \\ 1/ \Gamma _3 \end{bmatrix} $, and
\[
	[I _2] _{ (\sigma _3, \sigma _1, \sigma _0) } = \frac1{2 W _0} (\Gamma ^{-1} \tilde{P}) ^{-1} \begin{bmatrix} 1/ \Gamma _1 \\ 1/ \Gamma _2 \\ 1/ \Gamma _3 \end{bmatrix} 
		= \frac1{2 W _0} \tilde{P} ^{-1} \begin{bmatrix} 1 \\ 1 \\ 1 \end{bmatrix} \,. 
\]
Thus,
\[
	[I _2] _\beta = \tilde{P} \, [I _2] _{ (\sigma _3, \sigma _1, \sigma _0) } = \frac1{2 W _0} \begin{bmatrix} 1 \\ 1 \\ 1 \end{bmatrix} = s _0 = [\sigma _0] _\beta \,.
\] 
Therefore $ I _2 = \sigma _0 $.  It follows that $ I _2 (a _0, a _1, a _2, a _3) = a _0 $.
\end{proof}

\begin{remark}
	From the proof it is easy to see that, in the case $ W _0 < 0 $, choosing $ x \in \mathcal{S} $, $ Q(x) = 1 $, for the definition of Pauli symbols in \eqref{pauliSymbolsDefined}, amounts to interchanging the roles of $ a _1 $ and $ a _3 $ in the second expression of \eqref{shapeConics}.
\end{remark}

\section{Reduced space as coadjoint orbit}

The Poisson structure on $ \mathfrak{u}^\ast (2) $ induces a symplectic structure on its coadjoint orbits.  These can be identified with the symplectic reduced spaces of the 3-vortex problem when Heron's function takes the only physically meaningful value, zero.

\subsection{Coadjoint orbits}

As discussed in remark \ref{rem:angularImpulse}, $ M = - J $ when the center of circulation $ Z _0 $ is assumed to be at the origin, where $J$ is the momentum map of the $ SO(2) $ action.  Following common practice we will denote by $ \mu $ the value taken by the motion-invariant $J$, so $ a _0 = - \mu $.

We conclude that the symplectic leaves of $ \mathcal{V} $ are the submanifolds 
\[
	\mathcal{O} _{ (\mu, H) } \stackrel{\text{\tiny\rm def}}{=} \left\{ \sum _i a _i \sigma ^i \mid a _0 = - \mu \,, \; \signw(a _1 ^2 + a _2 ^2) + a _3 ^2 - \mu ^2 = H \right\} \,, 
\] 
with $ \signw = \sign(W _0) $.  Casimirs are constant on coadjoint orbits and, by a dimension count, it is easy to see that the $ \mathcal{O} _{ (\mu, H) } $ are precisely the coadjoint orbits of on $ \mathcal{V} \cong \mathfrak{u}^\ast $.

Physically meaningful dynamics occur only when $ H = 0 $.  Therefore, the symplectic reduced spaces for the three-vortex problem are the coadjoint orbits
\[
	\mathcal{O} _\mu \stackrel{\text{\tiny\rm def}}{=} \left\{ \sum _i a _i \sigma ^i \mid a _0 = - \mu \,, \; \signw(a _1 ^2 + a _2 ^2) + a _3 ^2 = \mu ^2 \right\} \,. 
\] 
In this way, given $ \mu \neq 0 $, $ \mathcal{O} _\mu $ is a 2-dim sphere of radius $ | \mu | $ (if $ W _0 > 0 $) or a two-sheeted 2-dim hyperboloid (if $ W _0 < 0 $).  In both cases, the conic ---sphere or hyperboloid--- sits inside the plane $ a _0 = - \mu $, so that the center fo the conic is on the $ a _0 $-axis.

The symplectic form on $ \mathcal{O} _\mu $ is $ 1/(2 \mu) $ times the area form induced by the metric $ d s ^2 $ on the ambient space $ \mathbb{R} ^3 $ (identified with the hyperplane $ a _0 = -\mu $).  This metric is Euclidean or hyperbolic depending on $ \sign(W _0) $; namely,
\[
	d s ^2 = \left\{ \begin{array}{r@{\quad\text{if}\quad}l}
		d a _1 ^2 + d a _2 ^2 + d a _3 ^2 & W _0 > 0 \,, \\[1ex]
		- d a _1 ^2 - d a _2 ^2 + d a _3 ^2 & W _0 < 0 \,. 
	\end{array} \right.
\] 
	
Since each point in the reduced space $ \mathcal{O} _\mu $ represents an equivalence class of vortex configurations with the same ``shape'', it is natural to call $ \mathcal{O} _\mu $ the \emph{shape-sphere} or -\emph{hyperboloid}.

\begin{remark}
	Below we will refer to the $ a _k $-axes, $ k = 1, 2, 3 $, giving a reference frame on the hyperplane $ a _0 = - \mu $, which is identified with $ \mathbb{R} ^3 $.  Thus, the origin (where the axes intersect) is at the point $ (-\mu, 0, 0, 0) \in \mathbb{R} ^4 $.
\end{remark}

\subsection{The area axis}

By definition of dual basis, each Pauli symbol $ \sigma _k $ is the projection functional giving the $ a _k $-component of a vortex configuration $ v \in \mathcal{V} $, i.e. $ \sigma _k (v) = a _k $.  This gives $ a _2 $ a simple interpretation in terms of the oriented area of the vortex triangle.  Indeed, from \eqref{vortexPauliCommRel},
\begin{equation} \label{areaPauliCoeff}
	a _2 = \frac{ 2 }{ \sqrt{| W _0 |} } \Delta \,, 
\end{equation}
where $ \Delta $ is the oriented area of the configuration.  In other words, the $ a _2 $-axis represents oriented area.  Hence the \emph{equator} $ a _2 = 0 $ corresponds to collinear configurations and the \emph{north} and \emph{south} hemispheres ($ a _2 > 0 $ and $ a _2 < 0 $, respectively) correspond to the two possible orientations of the triangle.  In the spherical case ($ W _0 > 0 $), the poles represent equilateral triangles with opposite orientation.

\begin{figure}
	\includegraphics{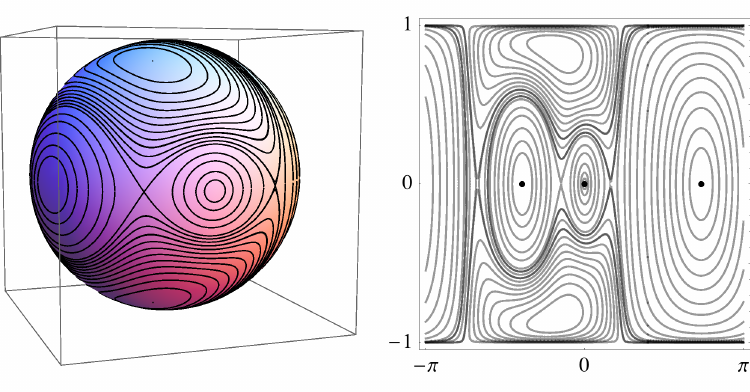}
	\hspace*{1.1in}(a)\hspace{2.49in}(b)
	\caption{\label{phaseportrait01}Phase portrait on $ \mathcal{O} _\mu $ and on its cylindrical coordinates chart for $ \Gamma _1 = 0.08904 $, $ \Gamma _2 = 0.28196 $, and $ \Gamma _3 = 0.629 $.  Solid dots on the cylindrical chart indicate binary collisions.}
\end{figure}

Figure \ref{phaseportrait01}-(a) shows, as an example, the phase portrait on $ \mathcal{O} _\mu $ for particular choices of the vortex strengths and $ \mu = 1 $.

\subsection{The $ a _1 $-$ a _3 $ plane}

Orienting the $ a _2 $-axis vertically, the three possible binary collisions represented by points on $ \mathcal{O} _\mu $ lie on the horizontal plane $ a _1 $--$ a _3 $.  The direction of those axes is determined by the choice of the $ x \in \tilde{\mathcal{S}} $ defining the Pauli symbols in \eqref{pauliSymbolsDefined}.  It is convenient to choose $ x \in \tilde{\mathcal{S}} $ so that a binary collision, say $ z _1 = z _2 $, lies on the $ a _3 $-axis.  With this choice the explicit expressions for Pauli symbols take their simplest form.  Also, a clear link with symplectic reduction using Jacobi coordinates can be established, as explained in the next section.  In what follows, assume $ \Gamma _1 + \Gamma _2 \neq 0 $ and choose $ \mu \in \mathbb{R} \setminus \{0\} $ so that $ \sign(\mu) = \sign(\Gamma _3 (\Gamma _1 + \Gamma _2) \Gamma _{\text{tot}}) $.

Let $ B _{ 12 } \in \mathcal{O} _\mu \subset \mathcal{V} $ be the binary collision corresponding to $ z _2 - z _1 = 0 $, so that $ | z _3 - z _1 | ^2 = | z _3 - z _2 | ^2 $.  That is to say, $ B _{ 12 } = \lambda ( \underline{\mathbf b} _1 + \underline{\mathbf b} _2 ) $ for some $ \lambda \in \mathbb{R} ^{+} $.  Using the coordinates of proposition \ref{prop:casimirsPauliCoords}, let $ B _{ 12 } = (a _0, a _1, a _2, a _3) $.  Requiring that $ B _{ 12 } $ be on the intersection of $ \mathcal{O} _\mu $ and the $ a _3 $-axis implies $ a _0 = \mu $, $ a _1 = a _2 = 0 $, $ a _3 ^2 = \mu ^2 $, and thus
\[
	\sigma _0 (B _{ 12 }) = \mu \,, \quad \sigma _1 (B _{ 12 }) = \sigma _2 (B _{ 12 }) = 0 \,, \quad \sigma _3 (B _{ 12 }) = \pm \mu \,.
\] 
The condition $ \sigma _1 (B _{ 12 }) = 0 $ means that $ \{ x, \overline{\mathbf \Delta} \} ( \underline{\mathbf b} _1 + \underline{\mathbf b} _2 ) = 0 $.  Expressed in terms of $ (x _1, x _2, x _3) := [x] _\beta $,
\begin{multline*}
	\Big( \big[ (\Gamma _3 - \Gamma _2) x _1 - (\Gamma _1 + \Gamma _3) x _2 + (\Gamma _1 + \Gamma _2) x _3 \big] \overline{\mathbf b} _1 \\
+ \big[ (\Gamma _2 + \Gamma _3) x _1 + (\Gamma _1 - \Gamma _3) x _2 - (\Gamma _1 + \Gamma _2) x _3 \big] \overline{\mathbf b} _2 \Big) ( \underline{\mathbf b} _1 + \underline{\mathbf b} _2 ) \\
= 2 \Gamma _3 (x _1 - x _2) = 0 \,. 
\end{multline*}
Together with \eqref{crossv0}, this implies that $ [x] _\beta $ is of the form $ (1, 1, - \frac{ \Gamma _{\text{tot}} + \Gamma _3 }{ \Gamma _{\text{tot}} - \Gamma _3 }) x _1 $.  More precisely,

\begin{proposition} \label{prop:PauliB12}
Let $ x \in \mathcal{W} ^\ast $ so that $ [x] _\beta = (1, 1, - \frac{ \Gamma _{\text{tot}} + \Gamma _3 }{ \Gamma _{\text{tot}} - \Gamma _3 }) / (4 W _0) $.
Then $ x \in \tilde{ \mathcal{S} } $ and the Pauli symbols defined by \eqref{pauliSymbolsDefined} are, explicitly,
\begin{align*}
	\sigma _1 &= \frac1{2 | W _0 | ^{ 1/2 } } \left( - \bar{b} _1 + \bar{b} _2 + \frac{ \Gamma _1 - \Gamma _2 }{ \Gamma _1 + \Gamma _2 } \, \bar{b} _3 \right) \,, \\[1ex]
	\sigma _2 &= \frac2{ | W _0 | ^{ 1/2 } } \bar{\Delta} \,, \\[1ex]
	\sigma _3 &= \frac1{2 W _0} \left( \frac{ \bar{b} _1 }{ \Gamma _1 }  +  \frac{ \bar{b} _2 }{ \Gamma _2  }  - \frac{ \Gamma _{\text{tot}} + \Gamma _3 }{ \Gamma _{\text{tot}} - \Gamma _3 } \, \frac{ \bar{b} _3 }{ \Gamma _3 } \right) \,.
\end{align*}
Moreover, $ B _{ 12 } = \lambda ( \bar{b} _1 + \bar{b} _2 ) $ with $ \lambda = 2 \Gamma _{\text{tot}} \, \mu / ((\Gamma _1 + \Gamma _2) \Gamma _3) $ and, expressed in $ a _k $-coordinates, $ B _{ 12 } = (\mu, 0, 0, \mu) $.
\end{proposition}

\begin{proof}
From proposition \ref{prop:qqM}, $ Q(x) = -1 $ iff $ x _1 = \pm 1/(4 W _0) $.  Direct substitution in \eqref{pauliSymbolsDefined} using \eqref{aaBeta} gives the stated expressions for the $ \sigma _k $, $ k = 1, 2, 3 $.  Let $ \mathbb{P} $ be the matrix changing $ (\sigma _0, \sigma _1, \sigma _3) $-coordinates to $ (\bar{b} _1, \bar{b} _2, \bar{b} _3) $-coordinates.  The columns of $ \mathbb{P} $ can be read off from the expressions for $ \sigma _1 $, $ \sigma _3 $ and $ \sigma _0 = (\sum _{ k = 1 } ^3 \bar{b} _k / \Gamma _k) / (2 W _0) $.  Then $ \mathbb{P} ^{ -T } $ transforms coordinates $ (a _0, a _1, a _3) $ to $ (b _1, b _2, b _3) $, and a direct computation shows that, with $ x _1 = \pm 1/(4 W _0) $, 
\[
	\mathbb{P} ^{ -T } \begin{bmatrix} \mu \\ 0 \\ \pm \mu \end{bmatrix} = 2 \mu \frac{ \Gamma _{\text{tot}} }{ (\Gamma _1 + \Gamma _2) \Gamma _3 } \begin{bmatrix} 1 \\ 1 \\ 0 \end{bmatrix} \,. 
\] 
For concreteness, choose $ x _1 = 1/(4 W _0) $.  Then $ a _3 = \mu $ and the claim follows.
\end{proof}

\begin{remark}
Our priviledged choice of Pauli symbols, given by proposition \ref{prop:PauliB12}, differs from the one given in \cite[eq. (5)]{BoLe98ii}.  With our choice the expression for the hamiltonian takes a simpler form.
\end{remark}

\section{Relation with symplectic reduction using Jacobi coordinates}
\label{sec:jbhCoordinates}

We want to relate the coordinates constructed in the previous section with the reduction obtained using Jacobi-Bertrand-Haretu (JBH) coordinates for three vortices.  (See \cite[\S 3.2]{Ma92} for a brief account of JBH coordinates.)

We consider the composition $ T _3 \circ T _2 \circ T _1 $ of three canonical transformations described below.  See \cite{HeSh2018} for a further discussion about these transformations including the verification of their canonical nature.

Let $ T _1 : \mathbb{C} ^3 \longrightarrow \mathbb{C} ^3 $, $ (z _1, z _2, z _3) \mapsto (Z _0, r, s) $ be given by
\begin{displaymath} %\label{firstCanTrans}
\begin{split}
	Z _0 &= \frac{ 1 }{ \Gamma_{\text{\rm tot}} } \sum _{ j = 1 } ^3 \Gamma _j \, z_j  \quad \text{(center of vorticity)} \,, \\
	r &= z _2 - z _1 \,, \\
	s &= z _3  -  \frac{ \Gamma _1 \, z _1 + \Gamma _2 \, z _2 }{ \Gamma _1 + \Gamma _2 }  \,. 
\end{split}
\end{displaymath}
Next, setting $ Z _0 = 0 $, let $ T _2 : \mathbb{C} ^2 \longrightarrow \mathbb{R} ^2 \times \mathbb{T} ^2 $ be given by
\begin{displaymath} %\label{jacobiVectors}
	r = \frac{ \sqrt{2 j _1} \, e ^{ \i \theta _1 } }{ \sqrt{A} }
	\,, \quad
	s = \frac{ \sqrt{ 2 j _2} \, e ^{ \i \theta _2 }}{ \sqrt{B} }
\end{displaymath}
with
\begin{displaymath} %\label{jacobiVectorsCoefficients}
	A := \frac{ \Gamma _1 \Gamma _2 }{ \Gamma _1 + \Gamma _2 }\,, \quad
	B := \frac{ (\Gamma _1 + \Gamma _2) \Gamma _3 }{ \Gamma _1 + \Gamma _2 + \Gamma _3 } \,.
\end{displaymath}
Finally, let $ T _3 : \mathbb{R} ^2 \times \mathbb{T} ^2 \longrightarrow \mathbb{R} ^2 \times \mathbb{T} ^2 $ be given by
\[
	I _1 = j _2 - j _1 \,, \quad I _2 = j _1 + j _2 \,, \quad \varphi _1 = \frac{ \theta _2 - \theta _1 }{ 2 } \,, \quad \varphi _2 = \frac{ \theta _1 + \theta _2 }{ 2 } \,. 
\] 
Then, as shown in \cite{HeSh2018}, $ (I _k, \varphi _k) $, $ k = 1, 2 $, are conjugate variables of the symplectic structure.

It easy to see that $ \varphi _2 $ keeps track of rigid rotations of the three-vortex configuration.  Since the hamiltonian is $ SO(2) $-invariant, $ \varphi _2 $ is a cyclic variable and $ I _2 $ is a constant of motion.  In fact $ I _2 $ is the angular impulse of the 3-vortex system (see remark \ref{rem:angularImpulse}), so $ I _2 = - \mu $, and the symplectic reduced space is parametrized by $ (I _1, \varphi _1) $ with symplectic form $ d I _1 \wedge d \varphi _1 $.  Indeed, a direct computation shows that, with $ (a _0, a _1, a _2, a _3) $ being the coordinates constructed in the previous section,
\begin{align}
	a _0 &= I _2 \nonumber \\
	a _3 &= I _1 \nonumber \\
	a _1 &= \sqrt{| I _2 ^2 - I _1 ^2 |} \, \cos(2 \varphi _1) \nonumber \\
	a _2 &= \sqrt{| I _2 ^2 - I _1 ^2 |} \, \sin(2 \varphi _1) \label{a3expression}
\end{align}

That is to say, $ (I _1, 2 \varphi _1) \in \mathbb{R} \times S ^1 $ are cylindrical coordinates of the shape-sphere or -hyperboloid $ \mathcal{O} _\mu $, with the cylindrical axis aligned with the $ a _3 $-axis.  (For the sphere, $ | I _1 | \le | \mu | $; for the hyperboloid, $ | I _1 | \ge | \mu | $.)

\begin{remark}
	Relation \eqref{areaPauliCoeff} can be recovered directly from \eqref{a3expression} and the geometric interpretation of the canonical transformation $ T _3 \circ T _2 \circ T _1 $.  Indeed:
	\[\begin{split}
		\Delta &= \frac12 |r| |s| \sin (\theta _2 -\theta _1) 
		= \sqrt{ \frac{ \Gamma_{\text{\rm tot}} }{ \Gamma _1 \Gamma _2 \Gamma _3 } } \, \sqrt{j _1 j _2} \, \sin(2 \varphi _1) \\
			&= \frac12 \sqrt{ \frac{ \Gamma_{\text{\rm tot}} }{ \Gamma _1 \Gamma _2 \Gamma _3 } } \, \sqrt{ I _2 ^2 - I _1 ^2 } \, \sin(2 \varphi _1)
			= \frac12 \sqrt{ \frac{ \Gamma_{\text{\rm tot}} }{ \Gamma _1 \Gamma _2 \Gamma _3 } } \, a _3 \,, 
	\end{split}\]
	which agrees with \eqref{areaPauliCoeff}.
\end{remark}

\subsection{Hamiltonian flow on $ \mathcal{O} _\mu $}

The Hamiltonian
\begin{equation} \label{vortexHamiltonian}
h = - \frac1{4 \pi} \sum _{ i = 1 } ^3 \Gamma _j \Gamma _k \ln b _i \,, \quad (i, j, k) \in C _3 \,, 
\end{equation}
induces a dynamic flow on the extended vortex configuration space $ \mathcal{V} $ which restricts to the reduced flow on the coadjoint orbits $ \mathcal{O} _\mu $, $ \mu \neq 0 $.  Since $ h $ does not depend on $ \Delta $, it follows that, for a fixed $ a _0 = \mu $, $h$ is a function of $ a _1 $ and $ a _3 $ only.  Thus, the level sets of $ h _\mu \stackrel{\text{\tiny def}}{=} \left. h \right|_{\mathcal{O}_\mu} $ on the shape conic $ \mathcal{O} _\mu $ are obtained by intersecting the cylinders 
\[
	C _{\mu, E} = \left\{ (a _1, a _2, a _3) \in \mathbb{R} ^3 \mid h(a _0, a _1, a _3) = E \,, \quad a _0 = \mu \right\}  
\]
with the sphere or hyperboloid $ \mathcal{O}_\mu $.  Now, since the cylindrical axis of $ C _{\mu, E} $ is along the $ a _2 $-direction, the phase portrait of the Hamiltonian flow will be most symmetrical when represented using cylindrical coordinates $ (a _2, 2\alpha) $ with respect to the $ a _2 $-axis.  For the compact case $ W _0 > 0 $ this means
\[
	a _3 = \sqrt{ \mu ^2 - a _2 ^2 } \, \cos 2\alpha \,, \quad a _1 = \sqrt{ \mu ^2 - a _2 ^2 } \, \sin 2\alpha \,.
\]
The coordinate transformation giving the cylindrical coordinates $ (a _2, 2\alpha) $ of $ \mathcal{O} _\mu $ in terms of $ (I _1, 2\varphi _1) $ is given by the equations
\[\begin{split}
	a _2 &= \sqrt{ \mu ^2 - I _1 ^2 } \, \sin(2 \varphi _1) \,, \\
	\tan 2\alpha &= \frac{ \sqrt{\mu ^2 - I _1 ^2 } }{ I _1 } \, \cos(2 \varphi _1) \,. 
\end{split}\]
It is easy to verify that $ d I _1 \wedge d \varphi _1 = d a _2 \wedge d \alpha $, so this coordinate transformation is cannonical.  (A similar coordinate transformation holds for the non-compact case $ W _0 < 0 $.)

As an example, figure \ref{phaseportrait01}-(b) shows the phase portrait on a cylindrical chart using coordinates $ (a _2, 2\alpha) $, for particular choices of the vortex strengths.

\section{Conclusions}

We described the reduced dynamics of a three-point-vortex system, with total vortex strength different from zero, as a Lie-Poisson system on coadjoint orbits of $ \mathfrak{u}^\ast (2) $ or $ \mathfrak{u}^\ast (1,1) $, corresponding to the compact and non-compact cases, respectively.  The classifying parameter discriminates these two cases is $ W _0 = \Gamma _{\text{tot}} / (\Gamma _1 \Gamma _2 \Gamma _3) $.  In both cases a basis of ``Pauli-symbols'' was constructed.  This means, in the compact case ($ W _0 > 0 $), that the elements of this basis satisfy commutation relations closely related to Pauli spin matrices.  Keeping one of the Pauli symbols proportional to the area functional, the construction of Pauli symbols was shown to depend on one parameter, $\theta$.  The Casimirs of the Lie-Poisson structure were shown to have very simple expressions in terms of a basis induced by the Pauli symbols.  These expressions were shown to be independent of $\theta$.  With their aid, the coadjoint orbits were easily identified with spheres or hyperboloids for the compact and non-compact cases, respectively.  Also, an explicit relation with symplectic reduction using Jacobi-Bertrand-Haretu coordinates was given.  In establishing this relation we found a priviledged choice of Pauli-symbols.

The methods used in this paper are very particular to the 3-vortex case.  Nevertheless, we belive that some ideas presented here may provide guidance in the study of other systems amenable to Lie-Poisson reduction.

%:Bibliography

\end{document}